\documentclass[preprint,superscriptaddress,aps,showpacs,showkeys]{revtex4}
\usepackage{amsfonts}
\usepackage{graphics}
\usepackage{graphicx}
\usepackage{slashed}
\newcommand{\be}{\begin{equation}}
\newcommand{\ee}{\end{equation}}
\newcommand{\en}{\end{equation}}
\newcommand{\ba}{\begin{eqnarray}}
\newcommand{\ea}{\end{eqnarray}}
\newcommand{\bea}{\begin{eqnarray}}
\newcommand{\eea}{\end{eqnarray}}
\newcommand{\bq}{\begin{eqnarray}}
\newcommand{\eq}{\end{eqnarray}}

\newcommand{\pa}{\partial}

\def\as{a\!\!\!/}

\def\ks{k\!\!\!/}

\def\ps{p\!\!\!/}

\def\bs{b\!\!\!/}

\def\Ds{D\!\!\!\!/}
\def\ds{\partial\!\!\!/}

\begin{document}

\title{On the anomalies in Lorentz-breaking theories}
\author{A. P. Ba\^eta Scarpelli}
\affiliation{Setor T\'{e}cnico-Cient\'{\i}fico, Departamento de Pol\'\i cia Federal,
Rua Hugo D'Antola, 95, Lapa, S\~{a}o Paulo, SP, Brazil}
\affiliation{Instituto de F\'{\i}sica, Universidade de S\~ao Paulo,
Caixa Postal 66318, 05315-970, S\~ao Paulo, SP, Brazi}
\email{scarpelli.apbs@dpf.gov.br}
\author{T. Mariz}
\affiliation{Instituto de F\'\i sica, Universidade Federal de Alagoas, 57072-270, Macei\'o, Alagoas, Brazil}
\email{tmariz@fis.ufal.br}
\author{J. R. Nascimento}
\affiliation{Departamento de F\'{\i}sica, Universidade Federal da Para\'{\i}ba\\
 Caixa Postal 5008, 58051-970, Jo\~ao Pessoa, Para\'{\i}ba, Brazil}
\email{jroberto, petrov@fisica.ufpb.br}
\author{A. Yu. Petrov}
\affiliation{Departamento de F\'{\i}sica, Universidade Federal da Para\'{\i}ba\\
 Caixa Postal 5008, 58051-970, Jo\~ao Pessoa, Para\'{\i}ba, Brazil}
\email{jroberto,petrov@fisica.ufpb.br}
\begin{abstract}
In this paper, we discuss the chiral anomaly in a Lorentz-breaking extension of QED which, besides the common terms that are present in the Standard Model Extension, includes some dimension-five nonminimal couplings. We find, using the Fujikawa formalism, that these nonminimal couplings induce new terms in the anomaly which depend on the Lorentz-violating parameters. Perturbative calculations are also carried out in order to investigate whether new ambiguous Carroll-Field-Jackiw terms are induced in the effective action.
\end{abstract}
\maketitle

\section{Introduction}

The topic of anomalies had a great importance in the history of quantum field theory and continues to be of intense interest nowadays (for a general introduction to this subject, see \cite{Fuji}). The anomaly, in its essence, is a violation of a classical conservation law by quantum corrections and, therefore, is entangled within the phenomenology of elementary particles. The paradigmatic example of the Adler-Bell-Jackiw anomaly \cite{ABJ,ABJ1}, also called chiral anomaly, attracted strong interest to other quantum symmetry violations, such as the conformal ones \cite{Fuji1,Fuji2}, which are very important in the gravitational context, being responsible for the trace anomaly of matter fields in a curved space-time \cite{AM}, and, further, gravitational anomalies \cite{Alv}. Besides these ones, essentially four-dimensional anomalies, one can also mention the parity anomaly \cite{Redlich}, typically occurring in odd-dimensional space-times (the original study has been carried out in three dimensions), the two-dimensional anomaly arising in the chiral Schwinger model \cite{JaRa,JackAmb} and the Callan-Harvey anomaly, relating contributions in two- and three-dimensional space-times \cite{CH,CH1}, further generalized for the noncommutative case \cite{CHNC}.

The interest in Lorentz-breaking models emerged with the seminal paper of Carroll, Field and Jackiw \cite{CFJ}. These theories are characterized by a widely extended list of possible couplings \cite{ColKost,ColKost1,PK} and, hence, of possible quantum corrections. This naturally raises some questions. First, how the presence of Lorentz-breaking terms modify the known anomalies? Second, may new anomalies, which have no analogues in the traditional Standard Model, emerge in the presence of Lorentz symmetry breaking? The subject of anomalies in Lorentz-breaking models has been studied before, like in the papers \cite{anomaly1} and \cite{anomaly2}, in which the chiral anomaly in the minimal Standard Model Extension (SME) \cite{ColKost,ColKost1} has been investigated, and no new corrections have been obtained. On the other hand, the absence of gauge anomalies in the SME have been shown in \cite{anomaly3}. Out of the scope of the minimal SME, Lorentz-breaking nonminimal couplings were considered in several papers within different contexts \cite{NM,NM1,NM2,NM3,aether,aether1,PQ,PQ1,axion,aether2,NMscarp,NMscarp1}, but these five-dimensional operators have not been considered in the study of quantum anomalies. In this paper, we consider the minimal SME with the addition of some five-dimensional nonminimal coupling terms in the study of the chiral anomaly. We show that these terms play an essential role in the appearance of new contributions to the chiral anomaly.

Furthermore, we study one-loop quantum corrections to the classical action. We particularly pay attention to corrections that could be associated with ambiguous coefficients which come from the cancellation of divergences, since they are similar to those involved in the perturbative calculation of the chiral anomaly. Indeed, it was argued in \cite{JackAmb} that the ambiguity of the CFJ term is related to the chiral anomaly, and it was shown in \cite{Chung,Chung1,Chung2,Chung3,Chung4} that it is actually related with the freedom in the definition of the conserved current.

The paper is organized as follows. In section II, we present the model and discuss some of its particularities; in section III, the formalism of Fujikawa \cite{Fuji} is used to calculate the chiral anomaly; section IV is dedicated to the calculation of one-loop corrections to the gauge sector and the investigation of its ambiguities; the features of the massless model are the subject of section V; and our conclusions are presented in section VI.

\section{The model}

Let us consider the following Lorentz-breaking extension of Quantum Electrodynamics (QED),
\bea
\label{act}
{\cal L}=\bar{\psi}(i\Gamma^m \tilde D_m-M)\psi-\frac{1}{4}F_{mn}F^{mn},
\eea
with
\bea
\label{gamma}
\Gamma^m=\gamma^m+c^{mn}\gamma_n+d^{mn}\gamma_n\gamma_5+e^m+f^m\gamma_5+\frac{1}{2}g^{mnl}\sigma_{nl},
\eea
\bea
\label{covariante}
\tilde D_m=\pa_m-i(eA_m+g_1a^nF_{nm}+g_2\epsilon_{mnlp}b^nF^{lp})
\eea
and
\bea
M=m+\as'+\bs\gamma_5+H^{mn}\sigma_{mn}.
\eea
Here, we use the $\Gamma^m$ and $M$, originally introduced in \cite{PK}, in this gauge-invariant Lagrangian density. We also included the nonminimal couplings responsible for the generation of the aether \cite{aether} and the axion \cite{PQ,PQ1,axion} terms. For simplicity, we suggest that the vector $b_m$ in the axial term $\bar{\psi}\bs\gamma_5\psi$ and in the magnetic coupling with coefficient $g_2$ is the same, just as in \cite{aether2}. These nonminimal interaction terms are dimension-five operators which are part of a more general class of terms of arbitrary dimensions, carefully studied in \cite{Kost-arb}. 

We will concentrate in the situations where ${a'}^m=d^{mn}=e^n=f^n=g^{mnl}=H^{nl}=0$, which encompass sufficiently interesting results to be discussed. We disregard all terms involving fourth- and higher-rank constant tensors.

In the next section, we concentrate in the calculation of the measure of the fermion path integral under a chiral transformation.

\section{The anomaly in the Fujikawa approach}

In the model we are considering, the $A_m$ is the usual gauge field, from which the field strength tensor,  $F_{mn}=\pa_mA_n-\pa_nA_m$, is constructed. As in the usual case, one can introduce the chiral current $j_m^5=e\bar{\psi}\gamma_m\gamma_5\psi$ and, after applying the chiral transformations $\psi(x)\to e^{ie\alpha(x)\gamma_5}\psi(x)$ and $\bar{\psi}\to \bar{\psi}e^{ie\alpha(x)\gamma_5}$, it is easy to see that at the classical level this current turns out to satisfy the relation
\bea
\label{id}
\pa_m(e\bar \psi\gamma^m\gamma_5\psi)=2ime\bar \psi\gamma_5\psi.
\eea
To verify this identity, one should consider the Lagrangian density of the extended spinor QED,
\bea
{\cal L}=\bar{\psi}(i \tilde \Ds-\bs \gamma_5- m)\psi,
\eea
for which the equations of motion are
\be
\left(i\gamma^m\pa_m+\gamma^m\tilde A_m-\bs \gamma_5 - m\right)\psi=0
\ee
and
\be
 \bar \psi\left(i \gamma^m \overleftarrow \pa_m -\gamma^m \tilde A_m + \bs \gamma_5 +m\right)=0,
\ee
with $\tilde A_m = eA_m+g_1a^nF_{nm}+g_2\epsilon_{mnlp}b^nF^{lp}$. Substituting these expressions in $\pa_m(e\bar \psi\gamma^m\gamma_5\psi)$, we immediately prove the relation (\ref{id}).

We note that our calculations represent themselves a generalization of known results through the replacement $eA_m\to eA_m+g_1a^nF_{nm}+g_2\epsilon_{mnlp}b^nF^{lp}$, since we restrict ourselves to the calculation of fermionic determinants, where the loop integrals are the same as in the usual case, and only the external legs are modified.
Then, one can repeat all the argumentation of \cite{Fuji} in the calculation of the measure of the functional integral in the fermionic field, replacing the standard covariant derivative with $\tilde D_m=\pa_m-i\tilde A_m$.

For the first step, we also set $c^{mn}=0$. Actually, the only difference from \cite{Fuji} in this case will consist in the fact that now, concerning the external gauge field, we replace $eA_m$ with $\tilde{A}_m$. Therefore, the commutator of the covariant derivatives will be given by $[\tilde D_a,\tilde D_b]=\tilde{F}_{ab}=\pa_a \tilde A_b-\pa_b \tilde A_a$, and one finds
\bea
\pa_m<e\bar \psi(x)\gamma^m\gamma_5\psi(x)>=2im<e\bar \psi\gamma_5\psi>+\frac{ie}{16\pi^2}\epsilon^{abcd}\tilde{F}_{ab}
\tilde{F}_{cd}.
\eea
This is the simplest Lorentz-breaking modification of the chiral anomaly, which clearly can involve higher derivative terms. The variation of the integral measure in this case will yield the following contribution to the effective action:
\bea
\label{standard}
\delta S_{measure}=\int d^4 x \frac{e}{32\pi^2}\alpha(x)\epsilon^{abcd}\tilde{F}_{ab}\tilde{F}_{cd},
\eea
where $\alpha(x)$ is the argument of the chiral transformation, cf. \cite{Chung,Chung1,Chung2,Chung3,Chung4}.

In order to simplify, we can obtain the result for particular conditions. If we set $g_2=0$ and get $\tilde{A}_n=eA_n+g_1a^mF_{mn}$, we have, from the Jacobian,
\bea
\delta S^{(1)}_{measure}&=&\int d^4 x \alpha(x)\left\{ \frac{e^3}{32\pi^2}\epsilon^{abcd}F_{ab}F_{cd}+\frac{e^2g_1}{8\pi^2}\epsilon^{abcd}F_{ab}a^n\pa_cF_{nd}+\right.\nonumber\\&+&\left.
\frac{eg_1^2}{8\pi^2}\epsilon^{abcd}
a^m\pa_aF_{mb}a^n\pa_cF_{nd}\right\}.
\label{anomaly1}
\eea

On the other hand, if we put $g_1=0$ to obtain $\tilde{A}_m=eA_m+g_2\epsilon_{mnlp}b^nF^{lp}$, we get the following result:
\bea
\delta S^{(2)}_{measure}&=&\int d^4 x \alpha(x)\left\{\frac{e^3}{32\pi^2}\epsilon^{abcd}F_{ab}F_{cd}+\frac{e^2g_2}{8\pi^2}\left[4b^m F_{mn} \pa_c F^{nc}
-b^c\pa_c \left(F_{nl}F^{nl}\right)\right]+ \right.\nonumber\\ &+&
\left.\frac{eg_2^2}{4\pi^2}\epsilon_{drql}b^r\left(\pa_c F^{ql}\right)\pa_a\left(b^a F^{cd}-b^c F^{ad}\right)\right\}.
\eea

It is interesting to carry out a discussion related with the consistency of the model. Here, we are treating a simplified theory with just one fermion. However, in a more realistic situation, this fermion should be included in a most complete theory like the Standard Model Extension (see f.e. \cite{ColKost,ColKost1}), in which it could appear as a chiral fermion. It is well known that in these situations, the axial anomaly leads to the violation of local symmetries, which in its turn spoil the unitarity of the theory. So, the phenomenological models should include a set of fermions in such a way that the anomaly would be cancelled out. In Abelian Lorentz invariant theories, for example, for a set of left-handed fermions, the cancellation occurs if they possess chiral charges, $Q_i$, such that $\sum Q_i^3=0$. In our expression for the anomaly in the case $g_2=0$, for example, we have terms which are even in the charge $e$. So, new conditions have to be imposed in the set of fermions which would compose the complete theory. Let us suppose that we have in the phenomenological theory a set of fermions with chiral charges $Q_i$ interacting nonminimally with the gauge field with coupling $G_i$. For the anomaly cancellation, they must be such that $\sum Q_i^3=\sum Q_i^2 G_i=\sum Q_i G_i^2=0$. One possible realization is a model with two fermions for which $Q_1=-Q_2$ and $G_1=-G_2$. So, in principle, one can obtain the cancellation of anomalies through corresponding extensions of the model (\ref{act}), by introducing additional fermions with different chiral charges, and repeating the calculations above for each of the fermions.

Finally, let us include the $c_{mn}$ parameter. Our classical Lagrangian will be written as
\bea
{ \cal L}=\bar{\psi}\left[i\left(\gamma^m+c^{mn}\gamma_n\right)\tilde D_m-M\right]\psi-\frac{1}{4}F_{mn}F^{mn}.
\eea
The best way to proceed in this case is as follows. We introduce a modified covariant derivative $\bar D_n=(\delta^m_n +c^m_n)\tilde D_m$ and a modified gauge field $\bar A_n=(\delta^m_n +c^m_n)\tilde A_m$, as in \cite{susyaether} (another procedure would consist in using modified Dirac matrices given by
$\bar{\gamma}^m=\gamma^m+c^{mn}\gamma_n$, like in \cite{ColMac}), where, for the sake of simplicity, we take $c_{mn}$ to be symmetric.
The variation of the measure will be obtained by the replacement of  $\tilde{F}_{mn}$ by $\bar F_{mn}=\bar{\pa}_m \bar A_n - \bar{\pa}_n \bar A_m$ in the standard result (\ref{standard}), and with the extra factor $\Delta$, where $\Delta={\rm det}^{-1}(\delta^m_a+c^m_a)$. The final result will be given by
\bea
\delta S_{measure}=\int d^4 x\frac{e}{32\pi^2}\alpha(x)\Delta\epsilon^{abcd}\bar{F}_{ab}\bar{F}_{cd},
\eea
which is exact in $c_{mn}$.
For small $c_{mn}$, it is found that $\Delta=1-c^m_m$, which is an example of the presence of a scalar generated by Lorentz-breaking within the expression for the anomaly (in the case $g_1=g_2=0$, i.e. $\tilde{A}_m=A_m$, an analogous calculation was performed in \cite{anomaly1}, where a similar result was obtained in terms of $A_m$, rather than $\bar{A}_m$).

\section{Ambiguities in the perturbative calculations}

In this section, we perform perturbative calculations in order to identify ambiguous radiative corrections with the form of the Carroll-Field-Jackiw (CFJ) term in the modified gauge field. These corrections would, in turn, decompose in new terms in function of the original gauge field with the same ambiguous coefficient. The CFJ term involves the gauge field and has the Chern-Simons form
\begin{equation}
S_{CFJ}=-\frac{1}{4}\int dx^{4}\varepsilon_{abmn}k_{AF}^{a}A^{b}F^{mn},  \label{CFJ}
\end{equation}
where $k_{AF}^{a}$ is a constant 4-vector that selects a space-time direction.

We consider one-loop corrections to the purely gauge sector represented by graphs as the one depicted in Fig. 1, where the dot denotes an insertion of the $\bs\gamma_5$ into the propagator.


\vspace*{1mm}

\begin{figure}[htbp]
\includegraphics{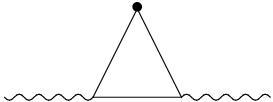}
\caption{Gauge field two-point function with one insertion of $\bs$.} \label{triangle}
\end{figure}

\vspace*{1mm}


Indeed, since we are searching for ambiguous corrections, we will consider here only one insertion of $\bs\gamma_5$. For two or more insertions the corrections are at most superficially logarithmically divergent, but, actually, the gauge invariant contributions should be superficially finite and, hence, unambiguous.

Using the same redefinition for the gauge field of the last section, beginning with $c_{mn}=0$, we reproduce the traditional result (cf. \cite{list,list1,list2,list3,list4,list5,ourLV1,ourLV2}):
\bea
\delta {\cal L}=C\epsilon^{abcd}b_a\tilde{A}_b\pa_c\tilde{A}_d,
\eea
where $C$ is an ambiguous constant coming from the cancellation of divergences and given by the integral \cite{ourLV1}
\bea
Cb_m=2i\int\frac{d^4l}{(2\pi)^4}\frac{b_m(l^2+3m^2)-4l_m(b\cdot l)}{(l^2-m^2)^3}.
\eea
Within different regularization schemes, $C$ is equal to $\frac{1}{4\pi^2},\frac{3}{16\pi^2},\frac{3}{8\pi^2}$, zero, etc. -- cf. \cite{list,list1,list2,list3,list4,list5,ourLV1,ourLV2}.

Let us again discuss some particular situations. The case with $g_2=0$ and $\tilde{A}_n=eA_n+g_1a^mF_{mn}$ needs some clarifications. We introduced the vector $a_m$ to get the correct transformation of the contributions to the action. In this case, one gets
\bea
\label{fireq}
\delta {\cal L}_1=C\epsilon^{abcd}(e^2b_aA_b\pa_cA_d+eg_1b_aa^nF_{nb}F_{cd}+g_1^2b_aa^ma^nF_{mb}\pa_cF_{nd}).
\eea
Alternatively, one can consider the case in which $b_m=0$ and  $M=m+\as\phi\gamma_5$, where $\phi$ is a pseudoscalar field slowly varying in space-time. In this case, we have the axion-like contribution
\bea
\label{fireq1}
\delta {\cal L}_{axion}=C\phi\epsilon^{abcd}(e^2a_aA_b\pa_cA_d+eg_1a_aa^nF_{nb}F_{cd}+g_1^2a_aa^ma^nF_{mb}\pa_cF_{nd}),
\eea
which is a higher-derivative generalization of the axion term obtained in \cite{axion}. Moreover, the ambiguity expressed by $C$ is the same of the chiral anomaly. In this sense, we obtained a relation between the induction of the axion term and the chiral anomaly for higher derivative terms.

For the case where $g_1=0$ and $\tilde{A}_m=eA_m+g_2\epsilon_{mnlp}b^nF^{lp}$, we get the result
\bea
\label{fireq2}
\delta {\cal L}&=&C\left\{e^2\epsilon^{abcd}b_a A_b\pa_cA_d+2eg_2 \left[b^2 F_{cd}F^{cd} -2 \left(b_c F^{cd}\right)^2\right]+ \right.\nonumber\\&+&
\left. 2g_2^2\left(b^2F^{cd}+b_a b^c F^{da}\right)\epsilon_{dmnp}b^m\pa_cF^{np}\right\},
\eea
in which the aether and the Myers-Pospelov \cite{MP} terms are induced.

Finally, we complete the calculation considering the $c_{mn}$ parameter, which we take as being symmetric. It is interesting that it is not necessary to do the calculations order by order in $c_{mn}$, since it is simple to obtain the exact result. We use a modified derivative, $\bar{\pa}_m=\pa_m+c_{mn}\pa^n$, and a modified gauge field, $\bar{A}_m=\tilde{A}_m+c_{mn}\tilde A^n$. Repeating all manipulations and introducing the ``twisted'' momentum $\bar{k}_m=k_m+c_{mn}k^n$, we see that the graph of Fig. 1 will yield the contribution
\bea
\delta {\cal L}=\bar{C}\epsilon^{abcd}b_a\bar{A}_b\bar{\pa}_c\bar{A}_d,
\eea
where the finite constant $\bar{C}$ is obtained from the expression
\bea
\bar{C}b_m=2i\int\frac{d^4l}{(2\pi)^4}\frac{b_m(\bar{l}^2+3m^2)-4\bar{l}_m(b\cdot \bar{l})}{(\bar{l}^2-m^2)^3}.
\eea
After the change in the integral measure, one finds that $\bar{C}=\Delta C$, where $\Delta={\rm det}\left|\frac{\partial l_a}{\partial\bar{l}_b}\right|={\rm det}^{-1}(\delta^m_a+c^m_a)$ is the Jacobian of the transformation from the ``old'' momentum $l^a$ to the ``new'' momentum $\bar{l}^a$.
So, our result for the correction is
\bea
\label{fireq3}
\delta {\cal L}=C\Delta\epsilon^{abcd}b_a\bar{A}_b\bar{\pa}_c\bar{A}_d.
\eea
Therefore, we succeeded in taking into account the complete one-loop impact of the $c_{mn}$.

A comment is in order now. The quantum induction of the CFJ term from the fermionic sector was the the subject of a lot of controversy. The main points in the discussions were the following ones: the dependence or not of the induced term on the regularization scheme; whether the vanishing of this term could be imposed by physical requirements, like gauge invariance, causality and unitarity; and, finally, whether the stringent limits on the magnitude of $k_{AF}^m$ impose restrictions on the existence of the CPT and Lorentz-breaking axial term in the fermionic sector.

For example, it was shown that this term with a space-like $k_{AF}^a$ vector would comply with microcausality and unitarity, whereas if it is time-like these requisites are spoiled \cite{Andrianov1}, \cite{Andrianov2}, \cite{Adam-Klinkhamer1}, \cite{Scarp-Belich}. Further, the issue of microcausality was studied in \cite{Adam-Klinkhamer2} from the quantum field theory viewpoint. It is known that the free fermionic sector \cite{Kost-Lehnert} can be maintained causal for a class of Lorentz-violating vectors, as well as the gauge part. However, when interactions are included and the effective action is calculated, the value of $C$ is independent of the particular $b_m$. For this reason, according to these papers, $C$ should be set to zero for general $b_m$.

From all these discussions, as summarized in \cite{P-Victoria},  we can state that the ambiguity of the radiative correction, $\Delta k_{AF}^m$, is
irrelevant for the physical content of the theory, since considering an effective constant
\begin{equation}
k_{AF}^{m \,eff}=(k_{AF}+\Delta k_{AF}+\delta k_{AF})^m,  \label{0.2}
\end{equation}
where $\delta k_{AF}^{m}$ is a finite counterterm for some normalization condition, one can always adjust the counterterm in order to obtain the
experimentally observed result. So, we believe these restrictions to the CFJ term do not contaminate the terms originally present in the fermionic Lagrangian density.

Finally, we would like to comment on another ambiguity which is present in the induced aether term discussed in \cite{aether,aether1}. In that case, where the unique coupling is the nonminimal one, the CPT-even aether term is generated by the ``fish'' graph of Fig. 2, which involves two nonminimal vertices.

\vspace*{1mm}

\begin{figure}[htbp]
\includegraphics{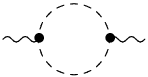}
\caption{Contribution with two nonminimal vertices.} \label{twonm}
\end{figure}

\vspace*{1mm}

It was argued in \cite{aether} that this graph arises in the theory whose fermion Lagrangian is
\bea
{\cal L}=\bar{\psi}(i\ds-g\epsilon^{abcd}b_aF_{bc}\gamma_d-m)\psi.
\eea
The induced term is given by
\bea
\Gamma_2=C_2g^2b^aF_{ab}b_cF^{cb},
\eea
where $C_2$ is an ambiguous constant equal, for example, to $\frac{1}{4\pi^2}$ or to 0.
Naively, one could expect that this ambiguity is also related to some anomaly. However, it is clear that this theory possesses only the so-called restricted gauge invariance. In other words, only the field $A_a$ suffers the usual gauge transformation, the spinor field remaining unchanged. At the same time, all argumentation in \cite{Fuji} is based on the chiral transformations $\psi\to e^{i\alpha(x)\gamma_5}\psi$. So, if $\alpha(x)=0$ (the spinor field is unchanged), there is no anomalous contribution. Therefore, this ambiguity is not related to the chiral anomaly, and the argumentation of \cite{JackAmb} cannot be applied here. Moreover, none of the known anomalies, including conformal one, seems to be related to this ambiguity.

\section{Considerations on the massless model}

There are very interesting particularities in the case where the model is massless. As we will show, it is easy to perform the one-loop calculation of the induced terms in all orders in the Lorentz-breaking parameters $b_m$, $a_m$, $c_{mn}$ and even in $d_{mn}$, which has not been considered in previous sections. First, let us write down our fermionic Lagrangian:
\be
{\cal L}_{\psi}=\bar \psi \left(i\Gamma^m \tilde D_m - \bs \gamma_5\right)\psi,
\ee
with
\be
\Gamma^m=\gamma^m +c^{mn} \gamma_n + d^{mn} \gamma_n \gamma_5
\ee
and $\tilde D_m$ given by eq. (\ref{covariante}). We use the redefinitions
\be
\bar A_n=\left(\delta^m_n+c^m_n\right)\tilde A_m,
\ee
\be
B_n=d^m_n A_n
\ee
and
\be
\tilde \pa_n=\left(\delta^m_n + c^m_n + d^m_n \gamma_5\right)\pa_m
\ee
to write
\be
\label{lpsi}
{\cal L}_\psi=\bar \psi\left[i\gamma^m\left(\tilde \pa_m -i \bar A_m -i \gamma_5 B_m\right)-\bs \gamma_5\right]\psi.
\ee
In terms of the redefined fields, the one loop correction to the purely gauge sector will be given by
\be
\Gamma_2(p)=-\frac 12 \bar A^m(-p)T^{VV}_{mn}\bar A^n(p) -\frac 12 B^m(-p)2T^{AV}_{mn}\bar A^n(p)
-\frac 12 B^m(-p)T^{AA}_{mn}B^n(p),
\ee
where the upper indices $V$ and $A$ are used to indicate, respectively, vectorial and axial vertices and
\be
T^{IJ}_{mn}=\int \frac{d^4k}{(2 \pi)^4} \mbox{tr} \left\{\Gamma^I_m S(k) \Gamma^J_n S(p+k)\right\},
\ee
with $\Gamma^V_m=\gamma_m$ and $\Gamma^A_m=\gamma_m \gamma_5$.

Let us calculate explicitly $T^{VV}_{mn}$. From the Lagrangian density (\ref{lpsi}), we write the propagator for the fermion field as
\be
S(p)=\frac{i}{\ps +c^{mn}\gamma_n p_m + \left(d^{mn}\gamma_n p_m -\bs\right)\gamma_5}.
\ee
This expression for the massless propagator is particularly interesting, since it allows for the decomposition
\be
S(k)=\frac{i}{\ks_1-\bs}P_L+\frac{i}{\ks_2+\bs}P_R,
\ee
where
\be
P_{R,L}=\frac{1 \pm \gamma_5}{2}
\ee
are the chiral projectors and
\be
k_{n1,2}=\left(\delta^m_n + c^m_n \pm d^m_n\right)p_m.
\ee
If we insert this decomposed propagator in the expression for $T^{VV}_{mn}$, we get
\bq
\label{tvv}
&&T^{VV}_{mn}=-\int \frac{d^4k}{(2 \pi)^4} \mbox{tr}\left\{ \gamma_m \frac{1}{\ks_1-\bs} \gamma_n \frac{1}{\ks_1+\ps-\bs}P_L\right\}- \nonumber\\ &&-
\int \frac{d^4k}{(2 \pi)^4} \mbox{tr}\left\{ \gamma_m \frac{1}{\ks_2+\bs} \gamma_n \frac{1}{\ks_2+\ps+\bs}P_R\right\} \nonumber \\
&&= -\Delta' \int \frac{d^4k}{(2 \pi)^4} \mbox{tr}\left\{ \gamma_m \frac{1}{\ks-\bs} \gamma_n \frac{1}{\ks+\ps-\bs}P_L\right\}
-\nonumber\\ &&- \Delta''\int \frac{d^4k}{(2 \pi)^4} \mbox{tr}\left\{ \gamma_m \frac{1}{\ks+\bs} \gamma_n \frac{1}{\ks+\ps+\bs}P_R\right\},
\eq
where $\Delta'= \mbox{det}^{-1} \left(\delta^m_n+c^m_n+d^m_n\right)$ and $\Delta''= \mbox{det}^{-1} \left(\delta^m_n+c^m_n-d^m_n\right)$ are the Jacobians for the changes in the momentum of integration, $k \to k_1$ and $k \to k_2$, respectively. As in the previous section, we are interested in CFJ-like terms. So, we have to select in eq. (\ref{tvv}) the terms with the $\gamma_5$ matrix inside the trace. It turns out that the only relevant term is
\be
T^{VV}_{mn5}=-\frac 12 (\Delta'+\Delta'')I_{mn5} ,
\ee
with
\be
I_{mn5}=\int \frac{d^4k}{(2 \pi)^4} \mbox{tr}\left\{ \gamma_m \frac{1}{\ks+\bs} \gamma_n \frac{1}{\ks+\ps+\bs}\gamma_5\right\},
\ee
where we used the fact that the integrals, due the appearance of the Levi-Civita tensor, are odd in $b_m$ and $p_m$. Similarly, we obtain
\be
T^{AV}_{mn5}=-\frac 12 (\Delta'-\Delta'') I_{mn5}
\ee
and $T^{AA}_{mn5}=T^{VV}_{mn5}$. It is easy to show that
\be
I^{mn}_5=-4iC'p_a b_b \epsilon^{mnab},
\ee
in which
\be
C'\eta_{mn}=\lim_{\mu^2 \to 0}\int \frac{d^4 k}{(2 \pi)^4} \left\{\frac{\eta_{mn}}{(k^2-\mu^2)^2}-4\frac{k_m k_n}{(k^2-\mu^2)^3}\right\}
\ee
is a regularization dependent surface term.

The constant $C'$ is strongly ambiguous. To illustrate this, we calculate it using two different procedures.

In the first scheme, we use the symmetry properties of the integral to replace $k_mk_n\to\frac{1}{4}\eta_{mn}k^2$. Afterwards, we do the Wick rotation, and we obtain
\bea
C'=\lim_{\mu^2 \to 0}\int \frac{d^4 k_E}{(2 \pi)^4}\frac{\mu^2}{(k^2_E+\mu^2)^3}=\frac{1}{32\pi^2},
\eea
which is the same both for $\mu^2\to 0$ and $\mu^2\neq 0$, such that the $\mu^2\to 0$ limit is well defined.

In the second scheme, we first promote the integral to $d$ dimensions, replace $\int\frac{d^4k}{(2\pi)^4}\to \kappa^{4-d}\int\frac{d^dk}{(2\pi)^d}$, being $\kappa$ an arbitrary parameter with mass dimension 1, and carry out the replacement $k_mk_n\to\frac{1}{d}\eta_{mn}k^2$. After the Wick rotation, we have
\bea
C'=\kappa^{4-d}\lim_{\mu^2 \to 0}\int \frac{d^d k_E}{(2 \pi)^4}\Big[\frac{1-\frac{4}{d}}{(k^2_E+\mu^2)^2}+\frac{4}{d}\frac{\mu^2}{(k^2_E+\mu^2)^3}\Big]=0,
\eea
which again is the same both for $\mu^2\to 0$ and $\mu^2\neq 0$, such that the $\mu^2\to 0$ limit is also well defined. We note that these calculations could be performed also with the use of the Schwinger proper time regularization.

It is interesting to compare our result with the ones of the section IV. First, let us consider that all Lorentz-violating parameters are small. So, we have $\Delta'=1-c^m_m-d^m_m$ and $\Delta''=1-c^m_m+d^m_m$, such that $\Delta'+\Delta''=2(1-2c^m_m)$ and
$\Delta'-\Delta''=-2d^m_m$. For $d^m_n=0$, the only contribution to the gauge sector comes from the $T^{VV}_{mn}$, since $B_m=0$, and the result is exactly the same obtained in the massive case. On the other hand, we see that if $d^m_n \neq 0$ the contribution to the CFJ term is of second order in $d^m_n$.

\section{Concluding comments}

We considered in this paper the generalization of chiral anomaly for Lorentz-breaking extensions of QED, which, besides the minimal SME terms, include some dimension-5 interaction operators. We have showed that new contributions from the measure of the fermionic functional integral, depending on the Lorentz-violating parameters, arise. The results have forms similar to the usual ABJ anomaly, written in terms of redefined gauge fields. Being written in function of the original gauge field, these forms decompose in new corrections to the divergence of the chiral current. As a consequence, chiral fermions in a phenomenological model have new conditions to be satisfied for the cancellation of the anomaly.

Perturbative calculations at one-loop order of corrections to the purely gauge sector have been performed. We focused on possible new ambiguous induced terms of the Carroll-Field-Jackiw form, which is known to be related with the chiral anomaly \cite{JackAmb}. Interesting new terms have been found, with some of them are dependent on scalar quantities constructed from the Lorentz-breaking parameters. We note that if the axial vector $b_m$ is zero, the new contributions to the anomaly simply do not arise neither from the measure nor from the triangle. Moreover, it is clear that one-loop functions with higher number of insertions  $\bs\gamma_5$ do not yield any anomalous contributions being superficially finite and hence unambiguous.

The open question is the possibility of anomalies in other Lorentz-breaking extensions. Therefore a natural continuation of this study could consist in the consideration of more sophisticated Lorentz-breaking extensions of QED, such as contributions involving higher-rank constant tensors. Such calculations naturally will involve higher derivatives, and the results for certain cases have been obtained in \cite{HD0,HD}. So, a natural perspective consists in the generalization of these results.

{\bf Acknowledgements.} This work was partially supported by Conselho
Nacional de Desenvolvimento Cient\'{\i}fico e Tecnol\'{o}gico (CNPq). The work by A. Yu. P. has been supported by the CNPq project No. 303783/2015-0.

\end{document}